\documentclass{aa}  
\usepackage{graphicx}
\usepackage{amssymb}

\usepackage{txfonts}
\begin{document}
   \title{A deep XMM-Newton X-ray observation of the Chamaeleon I dark cloud}

   \author{J. Robrade
          \and J.H.M.M. Schmitt
          }
   \offprints{J. Robrade}
   \institute{Universit\"at Hamburg, Hamburger Sternwarte, Gojenbergsweg 112, D-21029 Hamburg, Germany\\
       \email{jrobrade@hs.uni-hamburg.de} 
  }

   \date{Received 15 August 2006 / Accepted 20 September 2006}

 
  \abstract
 {Low-mass stars are known to exhibit strong X-ray emission during the early stages of evolution.
Nearby star forming regions are ideal targets to study the X-ray 
properties of pre-main sequence stars.}
   {A deep XMM-Newton exposure is used to investigate X-ray properties of the pre-main sequence population of the Chamaeleon I star forming region.}
  {The northern-eastern fringe of the Chameleon I dark cloud was observed with XMM-Newton, revisiting a region observed with ROSAT 15~years ago.
Centered on the extended X-ray source CHXR 49 we are able to resolve it into three major contributing components
and to analyse their spectral properties.
Furthermore, the deep exposure allows not only the detection of numerous, previously unknown X-ray sources, 
but also the investigation of variability and the study of the X-ray properties for the brighter targets in the field. 
We use EPIC spectra, to determine X-ray brightness, coronal temperatures and emission measures for these sources,
compare the properties of classical and weak-line T~Tauri stars
and make a comparison with results from the ROSAT observation.}
   {X-ray properties of T~Tauri stars in Cha I are presented. The XMM-Newton images resolve some previously blended
X-ray sources, confirm several possible ones and
detect many new X-ray targets, resulting in the most comprehensive list with 71 X-ray sources in the northern Cha~I dark cloud.
The analysis of medium resolution spectra shows
an overlapping distribution of spectral properties for classical and weak-line T~Tauri stars, with the X-ray brighter stars having 
hotter coronae and a higher $L_{\rm X}/L_{\rm bol}$ ratio. X-ray luminosity correlates with bolometric luminosity,
whereas the $L_{\rm X}/L_{\rm bol}$ ratio is slightly lower for the classical T~Tauri stars.
Large flares as well as a low iron and a high neon abundance are found in both types of T~Tauri stars.
Abundance pattern, plasma temperatures and emission measure distributions during quiescent phases are attributed to 
a high level of magnetic activity as the dominant source of their X-ray emission. 
   }
   {}


   \maketitle
%

\section{Introduction}

The Chamaeleon I dark cloud belongs to the Chamaeleon cloud complex, 
a well known star forming region at a distance of 
140\,--\,150\,pc. This compact region is rich in young stellar objects (YSO), and therefore ideal to investigate
low mass stars in early evolutionary phases. Typically optical H$\alpha$ emission 
and infra-red excess is used to classify pre-main sequence (PMS) stars and protostellar objects. 
The PMS stars with \hbox{M$_{*} \le 2 $\,M$_{\sun}$} are called T~Tauri stars, 
which are classified either as classical T~Tauri star (CTTS) 
or weak line T~Tauri star (WTTS).
While the strength of the H$\alpha$-emission is traditionally used to distinguish between classical (EW$>$ 10\,\AA) and weak line
T~Tauri stars, a classification based only on the time variable H$\alpha$-emission is often misleading.
Additionally, the lithium  absorption feature at 6707\AA\ is used 
to distinguishes PMS from field stars, since lithium is rapidly consumed in nuclear reactions and thus its absorption disappears with age.
Further, infra-red excess emission indicates
large amounts of circumstellar material, a typical selection criterion for very young stellar objects in general.
Chamaeleon I was surveyed extensively at optical and infra-red wavelength
for sources exhibiting one or several of the above characteristics
throughout the last decades, see e.g. \cite{gau92}, \cite{har93}, \cite{cam98} and \cite{gom01}.

X-ray observations showed a large number of additional sources in star forming regions that escaped previous detection.
The RASS (ROSAT All Sky Survey) data of the whole Chamaeleon region is presented by
\cite{alc95}, who report X-ray data and optical/near infra-red spectroscopy.
Dedicated observations of the Chamaeleon I dark cloud were performed with several missions and
its overall X-ray properties were already studied with {\it Einstein} IPC \cite{fei89}
and ROSAT PSPC \cite{fei93}.
Especially the two ROSAT pointings on the Chamaeleon I dark cloud performed in 1991 and presented in \cite{fei93}, 
revealed 70 X-ray sources (CHXR -- Chamaeleon X-ray ROSAT)
and 19 possible ones in the two fields, many of them unclassified.
The above authors  determined general X-ray properties
of the young stellar population in Cha I and already found a clear correlation between X-ray and bolometric luminosity.
However, these ROSAT data were only sufficient to determine positions and X-ray luminosities of these targets.
The X-ray detections triggered subsequent
optical follow up observations, see \cite{law96} who presented 
a HR diagram for the ROSAT detected PMS population of Chamaeleon I and references therein.
The new X-ray detected stars were found to be mainly WTTS, which in contrast to the CTTS lack
strong emission lines or infra-red excess. This is due to the absence of a significant disk and accretion onto the host star
in these usually more evolved stars. However, both types are commonly found in the same star forming region and show
an overlapping age-distribution. 
This is also the case for the Chamaeleon region with age estimations in the range of 1\,--10\,Myr by the above authors,
suggesting coevolution and individual timescales during their T~Tauri phase.

Both types of T~Tauri stars are strong and active X-ray sources with often very similar
X-ray properties. As in cool main sequence stars, magnetic activity is thought to be the origin
of most of its X-ray emission and variability. 
However, recent studies of high resolution X-ray spectra revealed accretion shocks as another
mechanism contributing to the X-ray emission in CTTS, see \cite{twc,twx,bptau,ctts}. 
However, this mechanism produces exclusively cool plasma and cannot be responsible for the bulk of X-rays seen in most TTS.
The {\it Chandra} Orion Ultradeep Project (COUP), a very deep ACIS observations of the Orion Nebula Cloud,
investigated the X-ray properties of its rich PMS population great detail (\cite{pre05}.
Magnetic activity is found to produce the bulk of their X-ray emission, whereas a strong $L_{\rm X}$/$L_{\rm bol}$
correlation is present. This is in contrast to cool main-sequence stars, where magnetic activity correlates with rotation,
indicating the presence of saturation and supersaturation effects in solar-type dynamos,
different internal structures, e.g. different convective turnover times or even the dominance of a different dynamo mechanisms.
Indications are found that accretion diminishes $L_{\rm X}$ on average and weakens the $L_{\rm X}$/$L_{\rm bol}$ correlation
for CTTS, where magnetic activity may also involve the circumstellar material, e.g. via star-disk interaction, but
the topology of their magnetic fields is virtually unknown.
First results of several shallower XMM-Newton pointings of the Cha I region are presented by \cite{tel06}.

In this paper we present a deep XMM-Newton observation
of the northern part of the Cha I cloud centered on the extended source CHXR\,49. 
The XMM-Newton field is fully included in the area covered by the northern ROSAT pointing and a comparison with
these measurements is possible for the whole field of view. 
All ROSAT X-ray sources are detected in the XMM-Newton exposure
as well as the possible detections. 
Furthermore, XMM-Newton spatially resolved several extended ROSAT sources
and additionally X-ray spectra could be obtained for these target. 
The large exposure time combined with XMM-Newton's sensitivity enables us to 
obtain X-ray spectra for these targets and to determine
spectral properties of the X-ray brighter sources in this region for the first time.
We use EPIC data to investigate the X-ray emission of these young
stars, determine their emission measure distribution and investigate possible spectral differences between CTTS and WTTS.
Additionally a large number of previously unknown fainter sources, often without known counterparts at other wavelengths, 
are discussed.
The outline of this paper is as follows: In Sect.\,\ref{obsana} we describe the observation and data analysis,
in Sect.\,\ref{res} we present the derived results subdivided into several topics, followed by a summary 
and our conclusions that are presented in Sect.\,\ref{sum}.

\section{Observation and data analysis}
\label{obsana}

The Cha I dark cloud was observed with XMM-Newton in a deep exposure, see Table\,\ref{obs} for an observation log. 
Data were taken with all X-ray detectors, which were operated simultaneously,
respectively the EPIC (European Photon Imaging Camera), consisting of the MOS and PN detectors
and the RGS (Reflection Grating Spectrometer).
Several intervals of high background due to proton flares are present in this data set, 
which were removed for spectral analysis and image creation. 
For the shown light curves only time periods of extremely high background have been removed, otherwise they are background subtracted.
Note, that while the light curves cover nearly the
complete exposure, additional time periods had to be discarded for spectral analysis, especially in the middle and toward the end
of the observation.
Due to overlapping photons from several sources in the RGS spectra and poorer signal to noise in these data,
we concentrate in this paper on the analysis of the EPIC data, which has sufficient data quality in all detectors.
 
\begin{table}[!ht]
\setlength\tabcolsep{4pt}
\caption{\label{obs}Observation log, duration total, (PN/MOS filtered).}
{
\begin{tabular}{lccc}\hline\hline
Target & Obs.Mode &  Obs. Time  & Dur. (ks)\\\hline
XX Cha & FF/med.\,F. & 2005-09-02T03:34--03/16:03 & 125 (70/85) \\
\end{tabular}
}
\end{table}

Data analysis was performed with the XMM-Newton Science Analysis System (SAS) software, version 6.5. Images, light
curves and spectra were produced with standard SAS tools and standard selection criteria were applied for filtering
the data \cite{sas}. 
Source detection was performed with the SAS-tool `edetectchain` independently with all EPIC detectors
and a visual cross-check of the results was carried out. 
X-ray spectral analysis was carried out with XSPEC V11.3 (\cite{xspec}) and
is performed in the energy band between 0.2\,--\,10.0\,keV, where data quality is sufficient 
and in the range 0.3\,--\,7.5\,keV for other sources respectively.
While the MOS detectors provide a slightly better spectral and spatial resolution, 
the PN detector is more sensitive and better suited especially for weaker sources.
The data of the detectors are analysed simultaneously but not co-added,
the background was taken from source free regions on each detector.
Modelling of the brighter targets performed with both PN and MOS data
uses a free normalization for each type of instrument to check for calibration uncertainties. 

For the analysis of the X-ray spectra we use multi-temperature models with three temperature components,
abundances are calculated relative to solar photospheric values as given by \cite{grsa}.
Such models assume the emission spectrum of a collisionally-ionized optically-thin gas
as calculated with the APEC code, see e.g. \cite{apec}.
For elements with overall low abundances and no significant features in the X-ray spectra,
i.e. Al, Ca, Ni, the abundances were tied to the iron abundance.
When data quality permits, we determine the abundances of individual elements, otherwise
typical global values, we adopt 0.5 solar abundances, are used.
We simultaneously modelled the temperatures and
mission measures (EM=$\int n_{e}n_{H}dV$) of the plasma components and checked the derived results for consistency.
X-ray luminosities were then calculated from the resulting best fit models.
Absorption in the circumstellar environment and in the interstellar medium is significant for 
most of our targets and is applied in our modelling as a free parameter.
In general the derived fit results are quite stable, but note that some of the fit parameters are mutually dependent.
Interdependence mainly affects the strength of absorption and emission measure of the cooler plasma at a few MK,
but also emission measure components and abundances
of elements with emission lines in the respective temperature range.
Consequently, models with different absolute values of these parameters but only marginal differences in its statistical
quality may be applied to describe the data, however ratios and relative changes of these properties are again a very robust result.
We checked other models, e.g. with a fixed temperature grid for the plasma components and overall consistent results were obtained.

Our fit procedure is based on $\chi^2$ minimization, therefore spectra
are always rebinned to satisfy the statistical demand of a minimum value of 15 counts per spectral bin.
All errors are statistical errors given by their 90\% confidence range and were calculated separately
for abundances and temperatures by allowing variations of normalizations and respective model parameters.
Note that additional uncertainties arise from uncertainties in the atomic data and
instrumental calibration which are not explicitly accounted for.


\section{Results}
\label{res}

The XMM-Newton observation covers a sky region of nearly 30\arcmin\,x\,30\arcmin\ including numerous X-ray sources. Due to the
different offset angles, detected counts and available spectral information about the nature of the sources,
we subdivide the sources into three groups of stars, which are
discussed below. The first section deals with the extended ROSAT source CHXR~49, followed by a sample of the other
CHXR sources covered by XMM-Newton. For these sources usually reliable classification from optical/IR spectroscopy
is available. In the last section we discuss properties of the further detected X-ray sources.
Stellar parameters given in this section are taken from \cite{fei93}, \cite{har93}, \cite{hue94} and \cite{law96}.

To determine the spectral properties of the stars, we extract spectra for each source from a
circular region whose size is adapted to the X-ray brightness of the target and its proximity to other stars.
While we cannot distinguish in the case of the often present absorption between extinction related
to circumstellar and cloud material, an extinction map of the Cha I cloud presented by \cite{cam97} and stellar values in \cite{cam98} 
indicate no strong cloud extinction for our field. For most areas cloud extinction is quite weak ($A_{V} \lesssim 0.5$), 
moderate cloud extinction ($A_{V}\sim 1-2$) is measured in the north-western and south-western area of our field.

\subsection{The central triple source CHXR 49}

\begin{figure}
   \centering
   \includegraphics[width=80mm]{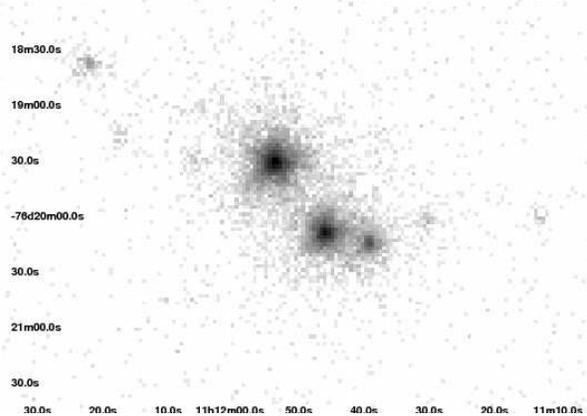}
      \caption{The MOS1 image of the extended ROSAT source CHXR49 reveals three main components, the stars Hn\,15, 
CHX\,18N and XX~Cha (upper left to lower right), surrounded by fainter sources.
               }
         \label{image1}
   \end{figure}

The MOS1 image of the center of the observed northern part of Cha~I star forming region is shown in Fig.\ref{image1}.
The extended ROSAT X-ray source, originally named CHXR~49, is clearly resolved by XMM-Newton into three major components, 
identified as Hn\,15 (now CHXR~49NE), CHX~18N and XX~Cha (CHXR~49) as 
suggested on the basis of optical finding charts by \cite{fei93}.
All three stars clearly show H$\alpha$ emission and infra-red excess (J-K\,$>$1) and their nature as YSO is quite certain.
While the M0 star Hn\,15 lies with a measured
H$\alpha$ EW=10\AA\, on the borderline between WTTS and CTTS, CHX\,18N with H$\alpha$ EW=8(3)\AA\, 
is of spectral type K1 and classified as WTTS 
and XX~Cha with H$\alpha$ EW=130\AA\, is a CTTS of spectral type M2. 
The determination of the spectral class for these variable objects is often not unique, 
given spectral types are typically mean values.

\begin{figure}
   \centering
   \includegraphics[width=75mm]{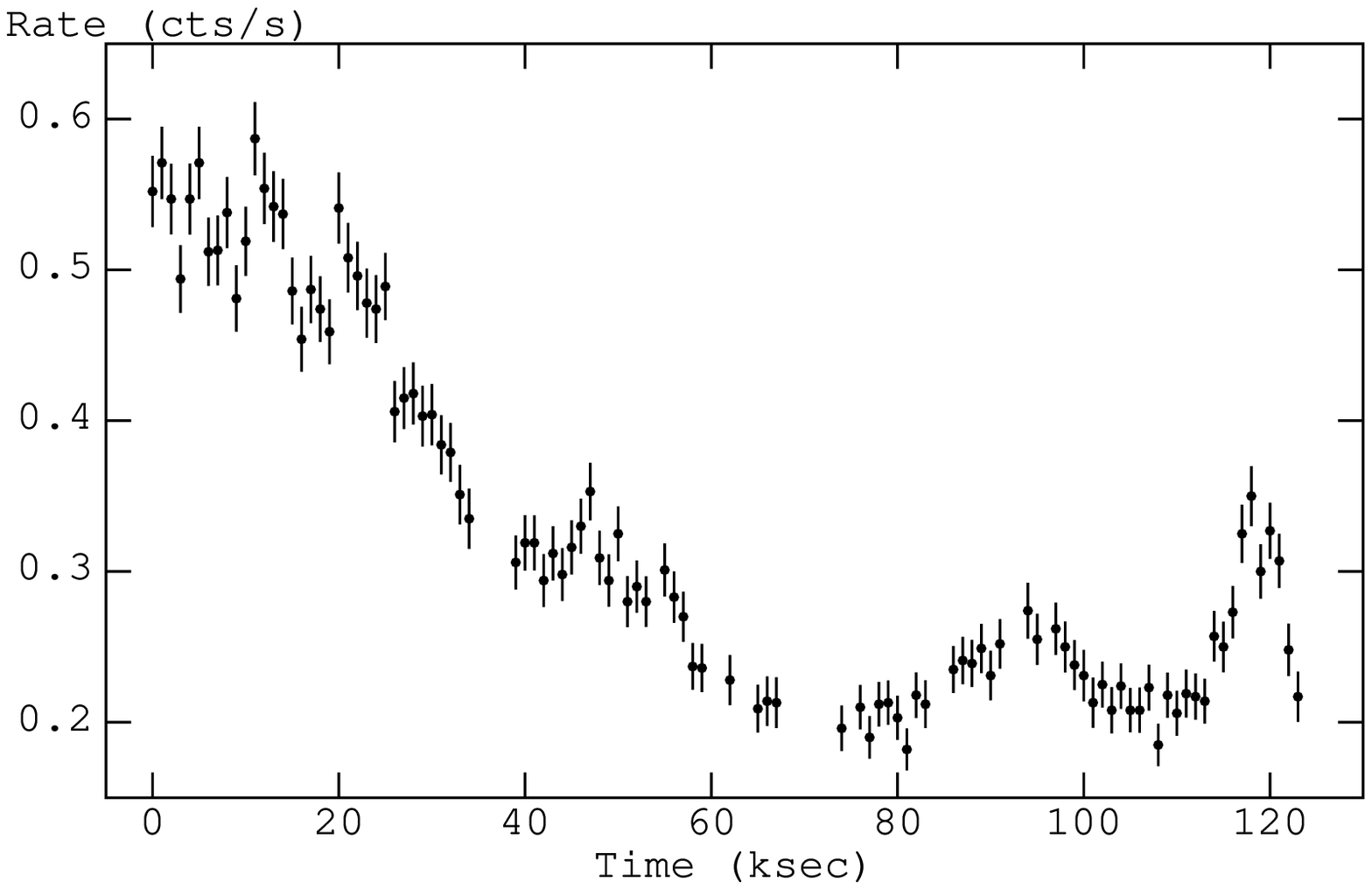}
   \includegraphics[width=75mm]{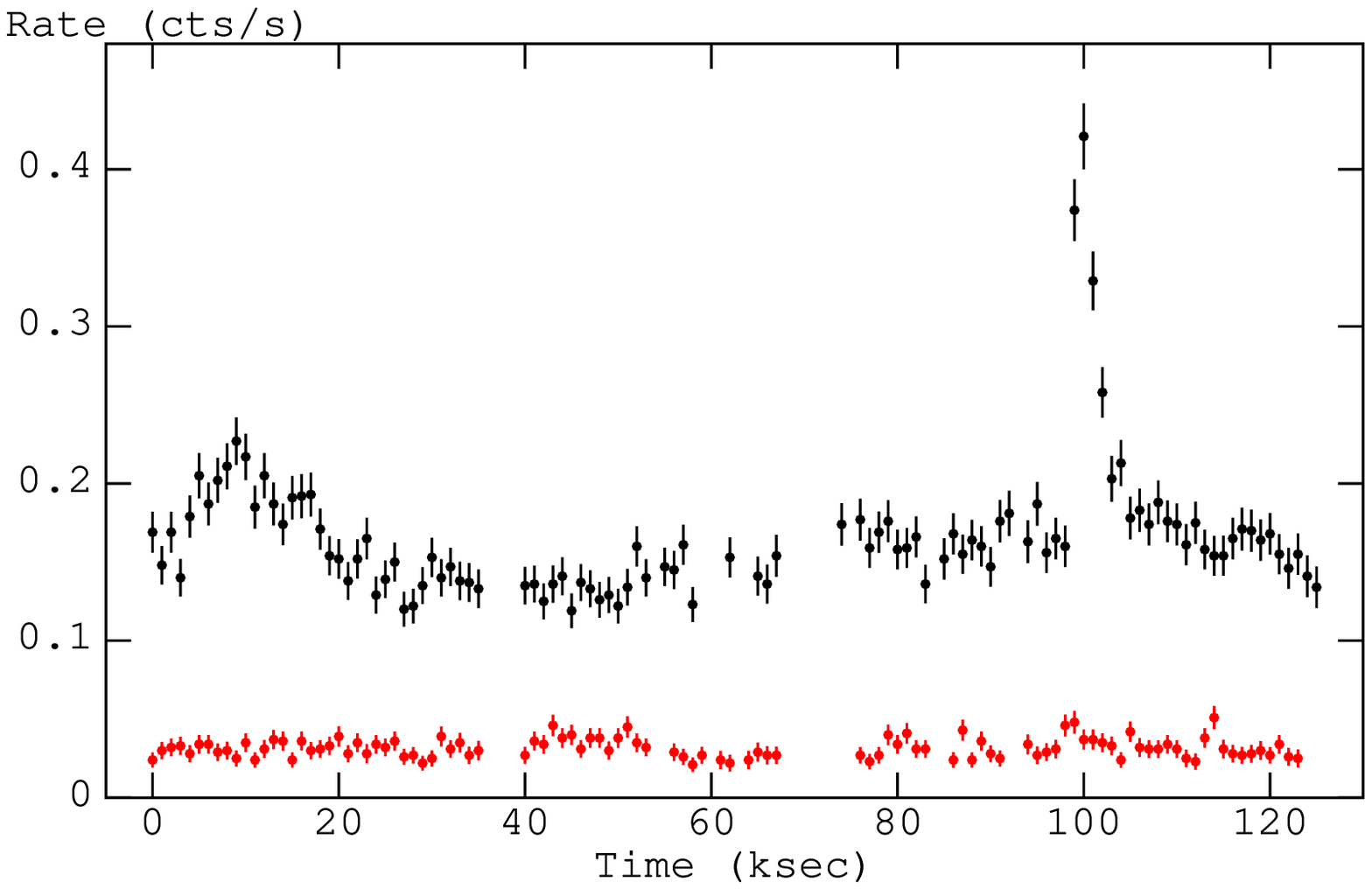}
      \caption{PN light curves of the three components of CHXR 49 with 1\,ks binning. 
Upper plot Hn\,15, lower plot CHX\,18N (black) and XX Cha(red/grey).}
         \label{lc_cen}
   \end{figure}

The light curves of the three components obtained from PN data are shown in Fig.\ref{lc_cen}; they are background subtracted and
periods with extreme background conditions have been discarded. 
Hn\,15 darkened by a factor of two to three over the first half of the observation,
probably a larger flare occurred some time before, and shows further smaller (factor $<$\,2) variability.
CHX~18N exhibits 
variable flux and a short impulsive flare with a duration of 5\,ks and an increase in count rate by a factor of roughly three. 
No significant variability was detected in XX~Cha over the whole observation. 
To investigate the stars at comparable conditions, we separately analyse the flare decay phase (h: 0--70\,ks) and the rest of the
observation for Hn\,15.

\begin{figure}
   \centering
   \includegraphics[width=50mm,angle=-90]{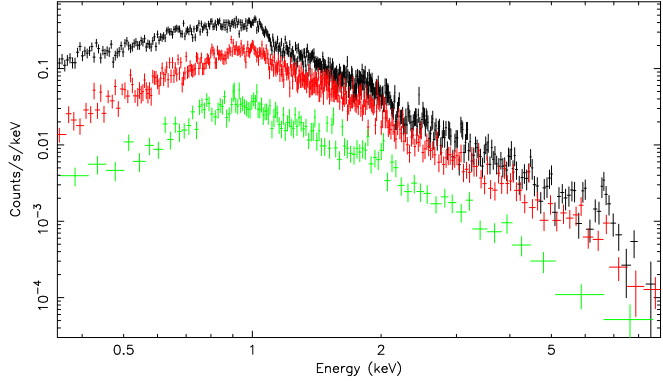}
      \caption{PN spectra of Hn\,15 (top), CHX18N (middle) and XX Cha (bottom) from the total observation.}
         \label{speccen}
   \end{figure}

Spectra are taken from a
circular regions covering the core of each star's PSF. The extraction radii, 25, 15, 7.5\arcsec\ for Hn\,15, CHX\,18N and XX~Cha 
respectively, are chosen to optimize the relation between the signal strength of the star and the 
contamination from the other sources. The spectra as measured by the PN detector are shown in Fig.\ref{speccen}.
Their shapes are quite similar,
only the weaker absorption and the contribution of hot flare plasma in the spectrum of Hn\,15 are striking.

The results of our spectral fits are given in Table\,\ref{spec1}, the quoted $L_{\rm X}$ is calculated from the best fit model 
in the energy range 0.2\,--10.0\,keV. 
Leaving aside the different X-ray luminosities and strength of absorption, the derived spectral properties of the target stars are comparable,
despite the values of their H$\alpha$ EW differing by over a magnitude. The X-ray spectra can be well described by a three temperature
model with plasma components at temperatures of around  2\,MK, 8\,MK and 25\,MK. The emission measure distributions (EMDs) are dominated by
the medium temperature and especially the hot plasma component at 25\,MK. Only for Hn\,15 in
the low state the medium temperature component is slightly stronger than the hot component.
The emission measure of the coolest component is quite weak in all cases, but poorly constrained due to the interdependence with
the absorption strength. 
Since only medium resolution EPIC spectra are available, we discuss only the elements which primarily contribute to the spectra.
Concerning the abundance pattern, a clear trend of higher abundances for high FIP elements is detected in all
stars as traced by the dominant elements iron, oxygen and neon. 
However, the other elements also fit into the picture of a metal depleted plasma at roughly 0.5 solar value with inverse FIP effect.
The low FIP element iron is always around 0.4\,--\,0.5 solar value, 
while the high FIP element neon is at solar value or above. Intermediate values are found for the medium FIP element oxygen, which is
always enhanced compared to iron but is found only in Hn\,15 around solar value.
Concerning the absolute values, we caution that other abundance tables 
or reduced emission measures would result in a higher average metallicity.

\begin{table}[!ht]
\centering
\caption{\label{spec1}Spectral X-ray properties, EPIC data ($N_{H}$ in $10^{21}$cm$^{-2}$, kT in\,keV, 
EM in $10^{52}$cm$^{-3}$ and $L_{\rm X}$ in $10^{30}$\,erg\,s$^{-1}$).}
{
\begin{tabular}{lccc}\hline\hline
Par. & Hn 15 (h)& CHX 18N & XX Cha \\\hline
$N_{H}$ & 1.2$^{+ 0.1}_{- 0.1}$ & 3.4$^{+ 0.3}_{- 0.2}$ & 3.9$^{+ 0.5}_{- 0.7}$ \\
kT1 & 0.10$^{+ 0.04}_{- 0.02}$ (0.12$^{+ 0.03}_{- 0.02}$) & 0.21$^{+ 0.04}_{- 0.02}$& 0.20$^{+ 0.06}_{- 0.07}$ \\
EM1 & 4.84$^{+ 6.30}_{- 1.28}$ (2.57$^{+ 6.16}_{- 1.50}$) & 5.11$^{+ 3.01}_{- 2.59}$ & 1.41$^{+ 1.04}_{- 0.49}$ \\
kT2 & 0.71$^{+ 0.03}_{- 0.03}$ (0.74$^{+ 0.02}_{- 0.02}$)& 0.71$^{+ 0.03}_{- 0.03}$ & 0.68$^{+ 0.05}_{- 0.06}$ \\
EM2 & 8.90$^{+ 0.77}_{- 0.70}$ (12.8$^{+ 0.90}_{- 2.80}$)& 9.27$^{+ 1.64}_{- 1.77}$ & 2.61$^{+ 0.32}_{- 0.31}$ \\
kT3 & 2.30$^{+ 0.30}_{- 0.22}$ (2.63$^{+ 0.16}_{- 0.13}$)& 2.11$^{+ 0.11}_{- 0.09}$ & 2.05$^{+ 0.20}_{- 0.17}$ \\
EM3 & 7.35$^{+ 0.96}_{- 0.94}$ (21.2$^{+ 1.10}_{- 1.10}$)& 14.6$^{+ 1.20}_{- 1.10}$ & 4.00$^{+ 0.37}_{- 0.43}$\\
Fe & 0.38$^{+ 0.06}_{- 0.06}$ & 0.40$^{+ 0.07}_{- 0.07}$ & 0.39$^{+ 0.18}_{- 0.14}$\\
O  & 1.03$^{+ 0.18}_{- 0.16}$ & 0.55$^{+ 0.21}_{- 0.14}$ & 0.49$^{+ 0.31}_{- 0.19}$\\
Ne & 1.14$^{+ 0.20}_{- 0.18}$ & 1.23$^{+ 0.27}_{- 0.22}$ & 1.32$^{+ 0.65}_{- 0.41}$\\\hline
$\chi^2${\tiny(d.o.f.)} & 1.09 (959) & 1.01 (633) & 0.89 (187)  \\\hline\hline
$L_{\rm X}$ obs. & 1.4 (3.1)& 1.4 & 0.35 \\
$L_{\rm X}$ emit.& 2.1 (4.7)& 3.4 & 1.0 \\\hline
\end{tabular}
}
\end{table}

When investigating spectral changes due to the variability we find for Hn\,15
the typical behaviour of a flare decay, i.e.
mainly the hot plasma component is affected, whereas especially the emission measure is varying. The 
cool component is fairly constant in temperature and EM, the medium temperature component is again fairly constant in temperature but decreases
in EM by roughly 50\%. The temperature of the hot component decreases from 30\,MK to 25\,MK, while its EM decreases by a factor of three.
The short and moderate flare on CHX\,18N and further minor variability on Hn\,15 is also accompanied by additional hot plasma 
as expected for magnetic activity but
have only minor effects on the derived spectral properties for the total data.

\subsection{Properties of ROSAT (CHXR) field sources}
\label{reschxr}

In this section we discuss the spectral properties of stellar X-ray sources 
and possible ROSAT detections as presented in \cite{fei93}. Regarding source detection,
all ROSAT sources within $\sim$\,15\arcmin\ around nominal pointing position are also detected by XMM-Newton, see Sect.\,\ref{futa}.
This includes the 'possible' X-ray sources CHXR~79, 80, 81 and 85, which are confirmed by our XMM-Newton observation.
The source CHXR~60 is clearly resolved as a double source with contributions from Hn\,18 and Hn\,19, 
where the NE component Hn\,19 is the X-ray brighter star.
The XMM-Newton positions coincidence well within errors with the given positions, identifications were first
presented in \cite{fei93} and \cite{law96}. For the comparison of the X-ray properties of CTTS and WTTS (Sect.\ref{specdiff})
we note that since H$\alpha$ emission is often highly variable, 
the discrimination between CTTS and WTTS is at least for stars with intermediate H$\alpha$ values arbitrary.
For these stars classification is not unique, e.g. CHXR~41 and CHXR~85 were classified as WTTS, 
whereas according to \cite{har93} they are clear CTTS.
The same holds again for the spectral classification, spectral types of these sources are 
mostly mid/late K to early/mid M, i.e. in the range K6\,--\,M4.
Several sources (CHXR~58, 80, 81) lack identified stellar counterparts and exhibit very hard spectra.
The hardest source CHXR~80 is additionally not present in the 2MASS/DENIS catalogues,
CHXR 58 and CHXR~81 exhibit IR counterparts of unknown classification (see Sect.\,\ref{futa}).

\begin{figure}
   \centering
   \includegraphics[width=76mm]{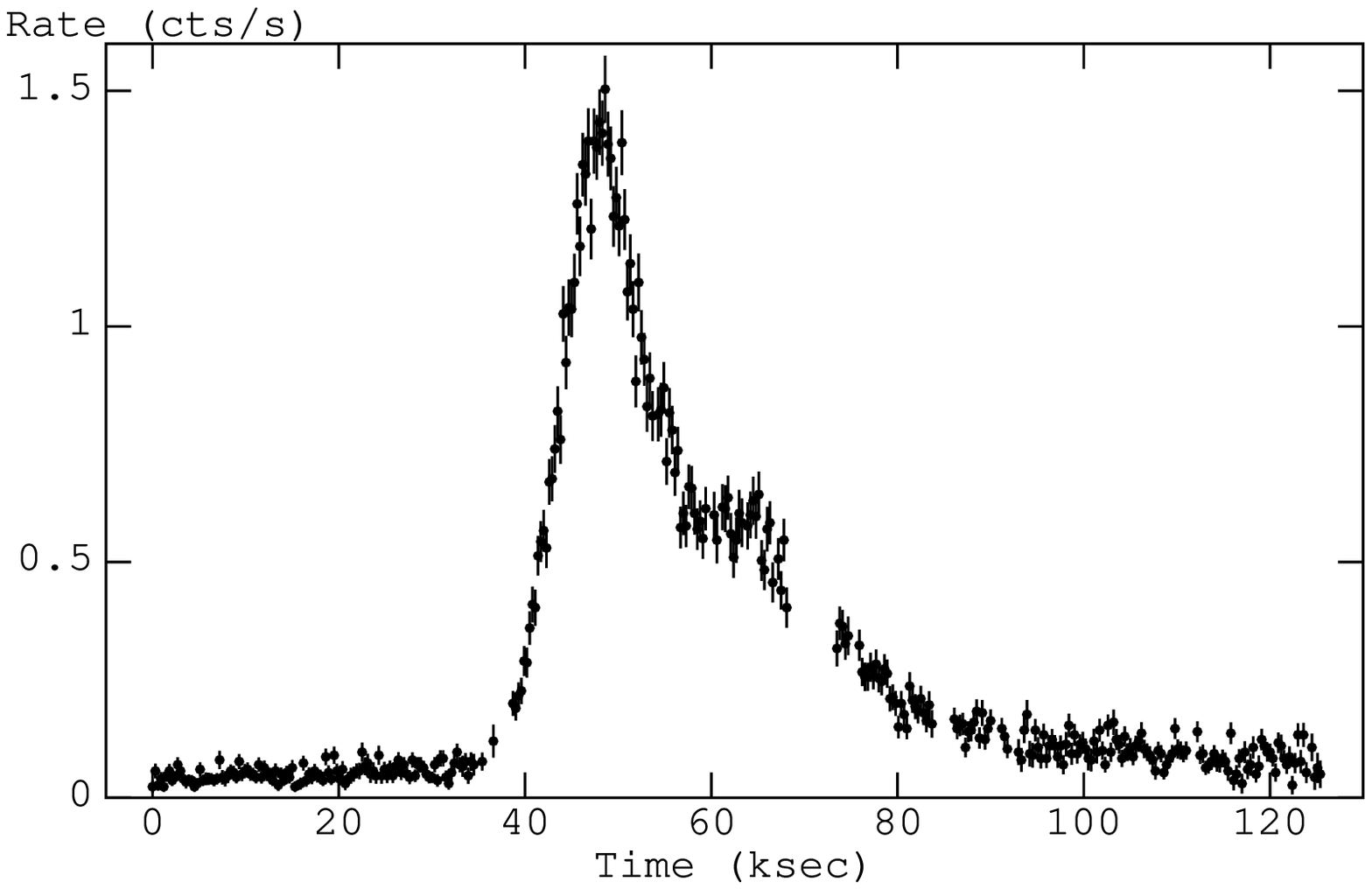}
 \includegraphics[width=76mm]{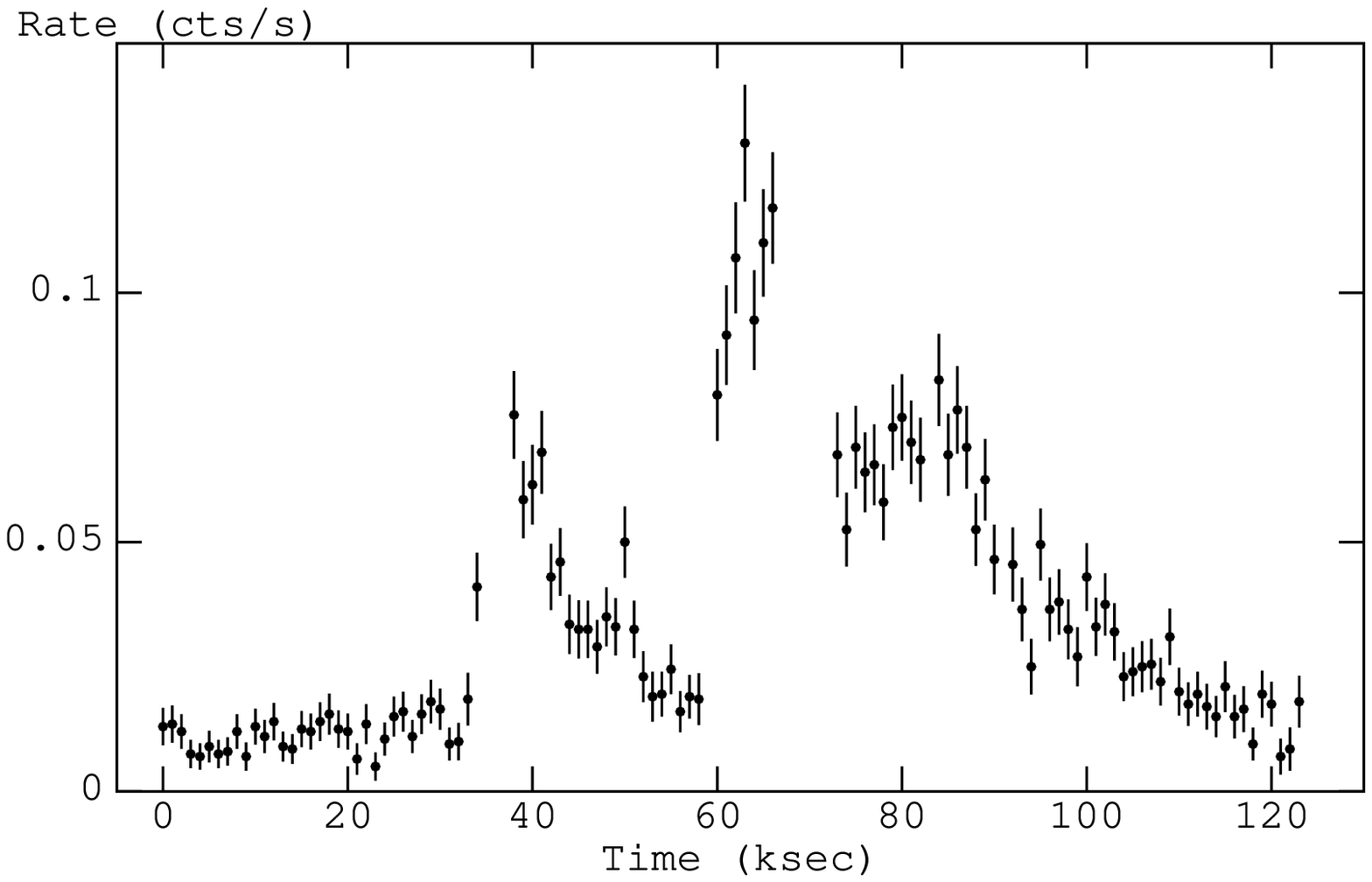}
      \caption{Larger flares observed on the WTTS CHXR 59 (top) and the CTTS CHXR 39 (bottom), PN light curve with 0.5/1.0\,ks binning. 
With a count rate increase of a factor of 30 the CHXR 59 event is the strongest flare observed in the field sources sample.}
         \label{lc59}
   \end{figure}

We first derived light curves of the sources and searched for variability and flares. Most stars show more or less constant
light curves, occasionally smaller activity and flares with variations within a factor of about two are observed.
In Fig.\,\ref{lc59} we show the light curves of the two stars where stronger variability is detected, i.e.
CHXR~39 with an active period including several flares (factor 5--8) and CHXR~59 with a flare of factor 30 in count rate
and an e-fold decay time of about 10\,--\,15\,ks. 
For spectral analysis we separate these datasets into phases of different activity to also allow an investigation of
their quiescent phase.
The observation of CHXR~59 is divided into a flare phase (h: 40--60\,ks), a 
decay phase (m: 60--85\,ks) and the remaining time (l: 0--40, 85--130\,ks) as quiescent phase, 
for CHXR~39 we define a quiescent phase (l: 0--30\,ks) and an active phase (h: 30--130\,ks).

 \begin{figure}
   \centering
   \includegraphics[width=50mm,angle=-90]{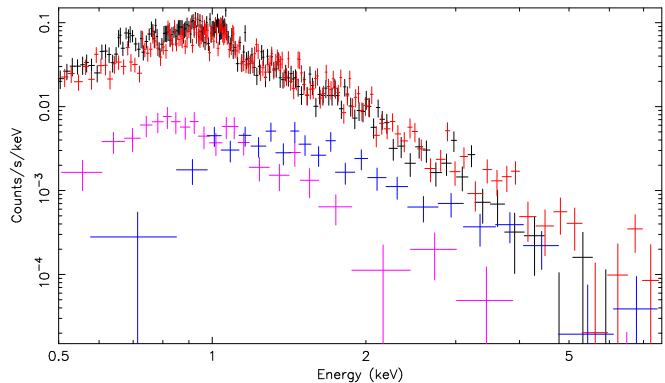}
      \caption{PN spectra of four TTS in the Cha I cloud, the WTTS CHXR~40 (black) and three CTTS, CHXR~46 (red),
35 (purple) and 79 (blue). }
         \label{specctts}
   \end{figure}

In Fig.\,\ref{specctts} we show the PN spectra of four known TTS without strong variability. 
All spectra of field sources are again extracted from circular
regions, whose size depends respectively on its X-ray brightness and proximity to other nearby sources. 
The spectra of the brighter targets CHXR~40 and 46, a clear WTTS/CTTS example, are quite similar, indicating the difficulty in
distinguishing TTS subtypes by medium resolution X-ray spectra.
Also shown are two examples of CTTS spectra with moderate data quality, including the heavily absorbed star CHXR~79.

The targets vary in data quality and detector coverage, for consistency we use
an absorbed three-temperature model with a global abundance value of 0.5 solar except neon at solar value for all sources.
The results of the spectral fits for these stars and the other CHXR sources are given in Table \ref{spec2}.
Not detected temperature components are indicated by dashes.
Typical errors on EM are as small as 10\% for the hotter components in the brighter sources, but note that
due to uncertainties in absorption and temperature of the cool components their emission measure may easily vary by a factor of two
for extreme cases already for the given absorption. We give the 90\% error interval of the cool EM under variation of the absorption in brackets
to indicate the large possible range, especially for data with low S/N.
Several other models, i.e models with variable abundances or with a cool component fixed at 0.2\,keV were tested,
and we find that the derived results agree well within errors.
The X-ray luminosity is given in the 0.2\,--\,10.0\,keV band and in brackets also the unabsorbed value, which was also used to calculate the
$L_{\rm X}/L_{\rm bol}$ ratio. While uncertainties in absorption affect the emitted X-ray luminosity,
the determination of $L_{\rm bol}$ is also a complex issue, especially for CTTS.
Here we used the values from \cite{law96} if available,
a comparison with otherwise used values given in \cite{che97} indicates an error on $L_{\rm bol}$ of up to 50\%.

The X-ray emitting level of our sample TTS scatters around the saturation-limit for magnetic activity at $L_{\rm X}/L_{\rm bol} \simeq 10^{-3}$
already during their low activity phases.
While the mean X-ray luminosity is with log $L_{\rm X}=29.9$\,erg/s nearly identical for the CTTS and WTTS subsample,
the mean log\,$L_{\rm X}/L_{\rm bol}$ ratio is \hbox{-3.3} for the CTTS and \hbox{-3.1} for the WTTS population.  
The log\,$L_{\rm X}/L_{\rm bol}$ ratio increases with X-ray luminosity, 
for the subsamples with log $L_{\rm X}$ values $>$\,30.0\,erg/s, 29.5\,--\,30.0\,erg/s and $<$\,29.5\,erg/s its value declines 
from \hbox{-2.95} over \hbox{-3.15} down to \hbox{-3.45}.
We identify two trends in the Cha I cloud TTS population, a lower $L_{\rm X}/L_{\rm bol}$ for CTTS compared to WTTS 
and an increase of average $L_{\rm X}/L_{\rm bol}$ 
with increasing $L_{\rm X}$. These trends are also present
in other star forming regions at different age, see e.g. \cite{pre05} for a large sample of overall younger stars in the ONC.

\begin{table*}[!ht]
\setlength\tabcolsep{5pt}
\centering
\caption{\label{spec2}Spectral X-ray properties of the CHXR sources as determined from XMM-Newton data. $N_{H}$ in $10^{21}$cm$^{-2}$, kT in\,keV, 
EM in $10^{52}$cm$^{-3}$ and $L_{\rm X}$ (0.2\,--10.0\,keV) in erg\,s$^{-1}$). 
The second last column names the used detectors and the number of spectral bins.}
{\scriptsize
\begin{tabular}{lccccccccccll}\hline\hline
Target & Class(H$\alpha$\,EW) & $N{_H}$ &kT1 & EM1& kT2 & EM2 & kT3 & EM3 & log $L_{\rm X}$ & log $L_{\rm X}/L_{\rm bol}$ &Det./bins & Other names\\\hline
CHXR35 & C (25.)& 1.5 & -- & -- & 0.62 & 0.48 & 1.70 & 0.38 & 28.8 (29.0)& -3.4 & PN/MOS2 (35) & Hn 8\\
CHXR37 & W (0.8) & 2.8 & 0.38 & 3.17 (1.42\,-\,4.95) & 0.84 & 2.32 & 1.72 & 5.41 & 29.8 (30.1)&  -3.2 & PN/MOS2 (246) & CCE98 1-74\\
CHXR39 (h)& C (71.)& 4.2 & 0.15 & 15.8 (7.3-120.)& 0.70 & 2.68 & 3.15 & 11.8 & 30.1 (30.5) & - & PN/MOS2 (252)& VZ Cha, Sz 31\\
CHXR39 (l)&  '' & 4.2 & 0.14 & 12.9 (5.5-57.6)& 0.38 & 2.77 & 2.02 & 2.29 & 29.4 (30.1) & -3.0 & PN/MOS2 (78) &\\
CHXR40 & W (1.7)& 2.2 & 0.39 & 3.45 (1.92\,-\,4.80) & 0.81 & 3.07 & 1.71 & 5.80 & 29.9 (30.2) & -3.1 &EPIC (383)& CHX 15b\\
CHXR41 & C (30.)& 6.3 & 0.24 & 4.78 (0.44-15.1)& 1.14 & 1.09 & 1.78 & 1.82 & 29.2 (29.9)& -3.3 &EPIC (79) & Sz 33\\
CHXR46 & C (56.)& 4.1 & 0.20 & 7.95 (3.81-29.7)& 0.72 & 6.82 & 1.78 & 7.30 & 29.9 (30.4) & -2.9 &EPIC (375)& WY Cha, Sz 36\\
CHXR53 & W (3.0) & 0.0 & 0.28 & 1.43 (1.11-1.81)& 0.71 & 1.24 & 1.36 & 1.27 & 29.7 (29.7)& -3.0 &EPIC (323) & CCE98\,1-109\\
CHXR59 (h)& W (3.0) & 0.8 & 0.32 & 22.6 (9.1-27.8) & 1.07 & 30.4 & 5.23 & 250. & 31.6 (31.7)& -&PN (394)&CCE98\,1-117\\
CHXR59 (m)& '' & 0.8 & 0.34 & 15.9 (13.3-19.1)& 1.67 & 30.6 & 5.10 & 35.8 & 30.9 (31.1) & -&PN (218)\\
CHXR59 (l)& '' & 0.8 & 0.22 & 3.15 (2.28-4.66)  & 0.78 & 3.45 & 2.17 & 10.4 & 30.2 (30.3) & -2.6 &PN (125)\\
CHXR60NE & W (1.5) & 1.0 & 0.34 & 0.66 (0.17-1.17)& 0.81 & 0.32 & 1.53 & 1.17 & 29.2 (29.4) & -3.1 &EPIC (88)& Hn 19\\
CHXR60SW & W (3.2) & 2.6 & 0.21 & 1.33 (0.0-2.60)& 0.58 & 0.62 & 1.20 & 0.72 & 29.0 (29.5) & -2.9 &EPIC (50) & Hn 18\\
CHXR62 & W (2.7) & 1.8 & 0.30 & 0.10 (0.0-2.49) & 0.66 & 0.66 & 1.50 & 0.34 & 28.9 (29.1)& -3.6 &EPIC (74)& Hn 20\\
CHXR79 & C (20.)& 9.2 & -- & -- & 1.08 & 1.42 & 2.51 & 1.96 & 29.2 (29.6) & -3.4 & PN (32)& Hn 9\\
CHXR85 & C (20.) & 7.0 & -- & --& 0.86 & 0.70 &--& --& 28.4 (29.0) & -3.7 &PN (10)& Sz 40\\\hline
CHXR58 & ? & 3.0 & 0.38 & 0.35 (0.0-1.29)& 2.02 & 2.93 & 8.62 & 2.33 & 29.7 & &EPIC (196)&\\
CHXR80 & ? & 3.3 & 0.30 & 0.18 (0.0-1.19)& 2.13 & 1.39 & $>$10.0 & 2.50 & 29.7 & &EPIC (146)&\\
CHXR81 & ? & 2.6 & 0.22 & 0.03 (0.1-0.46) & 0.43 & 0.19 & 6.10 & 1.27 & 29.2 & &EPIC (108)&\\\hline
\end{tabular}
}
\end{table*}

We note that the spectra of the 'hot' targets with kT3\,$>$\,5.0\,keV, i.e CHXR~58, 80, 81, 
may also be fitted at comparable quality with an absorbed thermal plasma components combined with a power-law.
However, the 6.7\,keV iron line complex which requires also hot thermal plasma is clearly present in CHXR~58 and 81, 
but not in the hottest target CHXR 80, indicating the possibility that these targets belong to two different types of X-ray sources.
No significant variability is present in the light curves of these targets,
therefore their spectral properties are not related to strong flares but have to be attributed to their quiescent state.
These sources exhibit harder spectra in comparison to the confirmed TTS members and
may be extremely hot, embedded YSOs or extragalactic sources, 
which could especially be the case for CHXR~80 since it also lacks an IR-counterpart.

Absorption towards known stellar targets is typically much higher for CTTS in our sample stars, the average $N{_H}$ 
is $5.4\times10^{21}$cm$^{-2}$ for the CTTS and $1.6\times10^{21}$cm$^{-2}$ for the WTTS population.
Only three WTTS have $N{_H}>2\times10^{21}$cm$^{-2}$
with two of them being clearly located in regions of higher cloud extinction. 
Absorption limits the conclusions that can be drawn about properties of the cool plasma, which is undoubtedly present.
No cool plasma component is detected in CHXR~35, 79 and 85, however these targets exhibit strong absorption and/or are faint.
Leaving aside the cool component, the observed T Tauri stars exhibit an overlapping distribution of EMDs.
Only three stars, two CTTS and on WTTS, have a hot plasma component at temperatures above 25\,MK (2.0\,keV) in quiescence.
Especially the fainter WTTS tend to show only moderately tempered coronae with average temperatures around 10\,MK and a hot component
at around 15\,MK. A correlation of the prominence of the hotter plasma and kT3 with $L_{\rm X}$ as is present in our sample,
especially in the WTTS population, but with quite low statistics. 

\begin{figure}
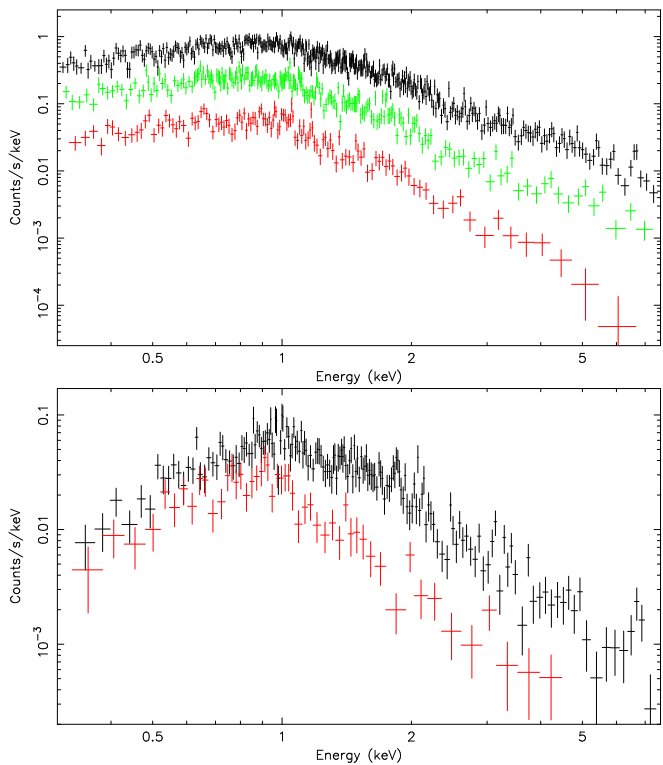

   \centering
   \includegraphics[width=50mm,angle=-90]{6250fi6a.ps}
   \includegraphics[width=50mm,angle=-90]{6250fi6b.ps}
      \caption{PN spectra of of the highly variable sources CHXR 59 (top) and CHXR 39 (bottom) during different phases of activity.}
         \label{specvar}
   \end{figure}

Spectral variability during flares and periods of stronger activity is investigated for CHXR~39 (CTTS) and CHXR~59 (WTTS), the spectra 
corresponding to the above defined phases of activity are shown in Fig.\,\ref{specvar}. Comparing the values of the respective 
low and high states, despite the X-ray brightening in CHXR~59 being a factor of five larger and the temperature of its
hot component with around 60\,MK nearly a factor two hotter than in CHXR~39, the spectral changes on CHXR~39 are comparatively 
harder. While the increase in the hot component is strongest in both cases, it is much more pronounced in CHXR~39
in comparison with the respective low state, which is in CHXR~59 already very hard.
Using the standard formula\,(14) for the loops half length
given by \cite{ser91} and putting in stellar parameters as well as the decay time of the large CHXR~59 event and its
peak temperature, we find a loop length of $\lesssim 0.3 R_{*}$. The upper limit results from the neglect of sustained heating.
These loops sizes and temperatures are
also typical for larger flares on very young active main sequence stars as expected for a WTTS, which lack a magnetically connected disk.

\subsection{Long term variations - XMM-Newton vs. ROSAT}

\begin{figure}
   \centering
   \includegraphics[width=90mm]{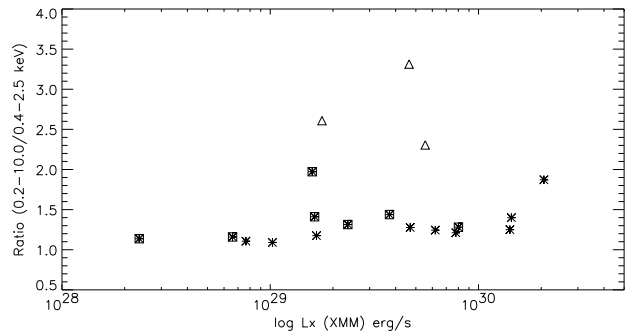}
      \caption{Comparison of $L_{\rm X}$-ratio in the XMM and ROSAT band vs. $L_{\rm X}$ (0.2\,--10.0\,keV) for the quiescent phase of the 
identified stellar targets, i.e. WTTS (asterisk) and CTTS (asterisk with box), as well as unknown sources (triangles).}
         \label{lxcomp}
   \end{figure}

For a comparison of the X-ray brightness during the XMM-Newton observation with ROSAT measurements performed in 1991, 
we also calculated the X-ray luminosity in the ROSAT PSPC band (0.4--2.5\,keV) from our models
and compare the values with the ROSAT measurements as given in \cite{law96}. The scaling factor of $L_{\rm X}$ for our targets between
the XMM-Newton (0.2\,--10.\,keV) and ROSAT PSPC energy band is shown in Fig.\ref{lxcomp}. 
For the quiescent phase the identified stellar sources occupy a strip in the range 1.1--1.4. 
The present trend of the X-ray brighter targets with a
higher ratio is due to their hotter coronae, while most of the scatter is caused by the different absorption towards the stars.
One outlier is the hard and heavily absorbed CTTS CHXR~79, the other the active source CHXR 59, both with a ratio of 1.9. 
The unidentified sources are above all other targets due to their strong emission at higher energies.

Within a sample of 13 stellar targets, we find an agreement between the measured X-ray luminosities from XMM-Newton and ROSAT
within 50\% for 6 and within factor two for 11 sources, whereas brighter and fainter
targets are distributed almost equally. These variations can be attributed to minor activity that is present the data
and the simple count rate flux conversion used for the ROSAT values.
Therefore we conclude that large long-term variations are not present in these stars and
no strong cross-calibration bias exists. Deviations exceeding factor two are present in two sources.
First, in flaring target CHXR~59 even the
low state during our observations is roughly a factor four brighter than the ROSAT value and second in the heavily absorbed, hot target CHXR~79
where the simple scaling factor used for the ROSAT data significantly underpredicts its X-ray luminosity.

\subsection{Spectral differences - CTTS vs. WTTS}
\label{specdiff}

In this section we investigate spectral parameters suitable to distinguish CTTS from WTTS via X-ray spectroscopy.
As already mentioned, the origin of the X-ray emission of WTTS is attributed to coronal emission in analogy to main sequence stars.
in CTTS part of the X-ray emission may also be produced via magnetically funneled accretion.
Usually density sensitive lines, e.g. in the \ion{O}{vii}-triplet, are used to detect high density plasma
that is attributed to strong accretion shocks. Cool high density plasma is used as a tracer of accretional X-rays,
see e.g. \cite{ctts} for examples of the analysis of high resolution X-ray spectra from CTTS
with XMM-Newton data. TW~Hya is the prototype of an accretion dominated star with an EMD mainly composed of
large amounts of cool high density plasma with temperatures around 3\,MK. It also exhibits an extreme abundance pattern
with Ne/Fe or Ne/O ratios in the range of ten as shown by \cite{twx}. However, TW~Hya is an exceptional case that is
coincidentally much closer than the well known star forming regions and its absorption is
negligible due to a nearly pole-on view. 
A probably more common scenario for a  CTTS is a star dominated by coronal activity and an additional contribution of
accretional plasma as seen for the first time in BP~Tau in \cite{bptau}.
Two additional CTTS observations were recently presented, enlarging the sample of high-resolution spectra. The close binary CTTS V\,4046 Sgr clearly shows
low f/i-ratios in the He-like triplets of \ion{O}{vii} and  \ion{Ne}{ix}, indicating high density plasma (\cite{v4046}.
Contrary, the spectrum of the CTTS system T~Tau shows a more typical coronal f/i-ratio in the \ion{O}{vii}-triplet (\cite{gue06}.
This suggests that accretion does not invariably produces high density plasma emitting at X-ray wavelengths, which might depend on the
respective accretion filling factor, or that a strong coronal contribution is present.

Here we present a sample of medium resolution spectra of known WTTS and CTTS and
compare the properties of members of both types of T~Tauri stars. Not surprisingly the CTTS show on average three times
stronger X-ray absorption compared to the WTTS due to larger amounts of additional circumstellar material. 
For example, the strong absorption seen in the spectra of CHXR~41 and 85 supports their CTTS nature.
However, both distributions significantly overlap and cloud absorption as well as viewing angle effects are surely present,
therefore absorption alone is no strong criterion.
Since accretional processes generate cool plasma
with temperatures around 3\,MK, plasma temperatures and EMDs may be used to distinguish the two TTS populations.
However, cool plasma can certainly be also of coronal origin, therefore
the pure existence of large amounts of cool plasma is again not a unique indicator for accretion.
On the other hand, young stars are known to be quite active and the presence of large amounts of cool plasma combined with the
absence of large amounts of hotter plasma may be a used as a tracer for an accretion dominated star.
Admittedly, a star that is dominated by cool plasma like TW~Hya is hard to identify in the
presence of strong absorption which diminishes the flux at the wavelength of interest.
CHXR~39 in its low state is such an example of a star with a strong cool plasma component and much smaller emission measure
at medium and high temperatures, but it is not as extreme as TW~Hya and much stronger absorbed, so a definite conclusion cannot be derived.
At higher temperatures the EMDs are dominated by coronal plasma and activity, independent of their WTTS or CTTS nature. 
For our sample stars, dominating medium and hot temperature plasma is the much more commonly found scenario.
For the abundance pattern again no clear distinguishing criterion exists, here the inverse FIP effect or grain depletion in the
accreted plasma may produce a comparable abundance pattern. An extreme Ne/Fe or Ne/O ratio in the order of ten
is seen in none of our sample stars.

In summary, a target like TW~Hya with its outstanding properties may be identified by medium resolution spectroscopy. However, the
supposably more common sources like BP~Tau with stronger absorption and dominant coronal contributions to the X-ray emitting 
plasma require deep exposures and high resolution spectroscopy to distinguish possible X-ray generating mechanisms.
While an accretional contribution in some of the CTTS may be present, clearly most of our sample targets are strongly dominated by
X-ray emission originating from magnetic activity.

\subsection{New X-ray sources}
\label{futa}

\begin{figure}
   \centering
 \includegraphics[width=85mm]{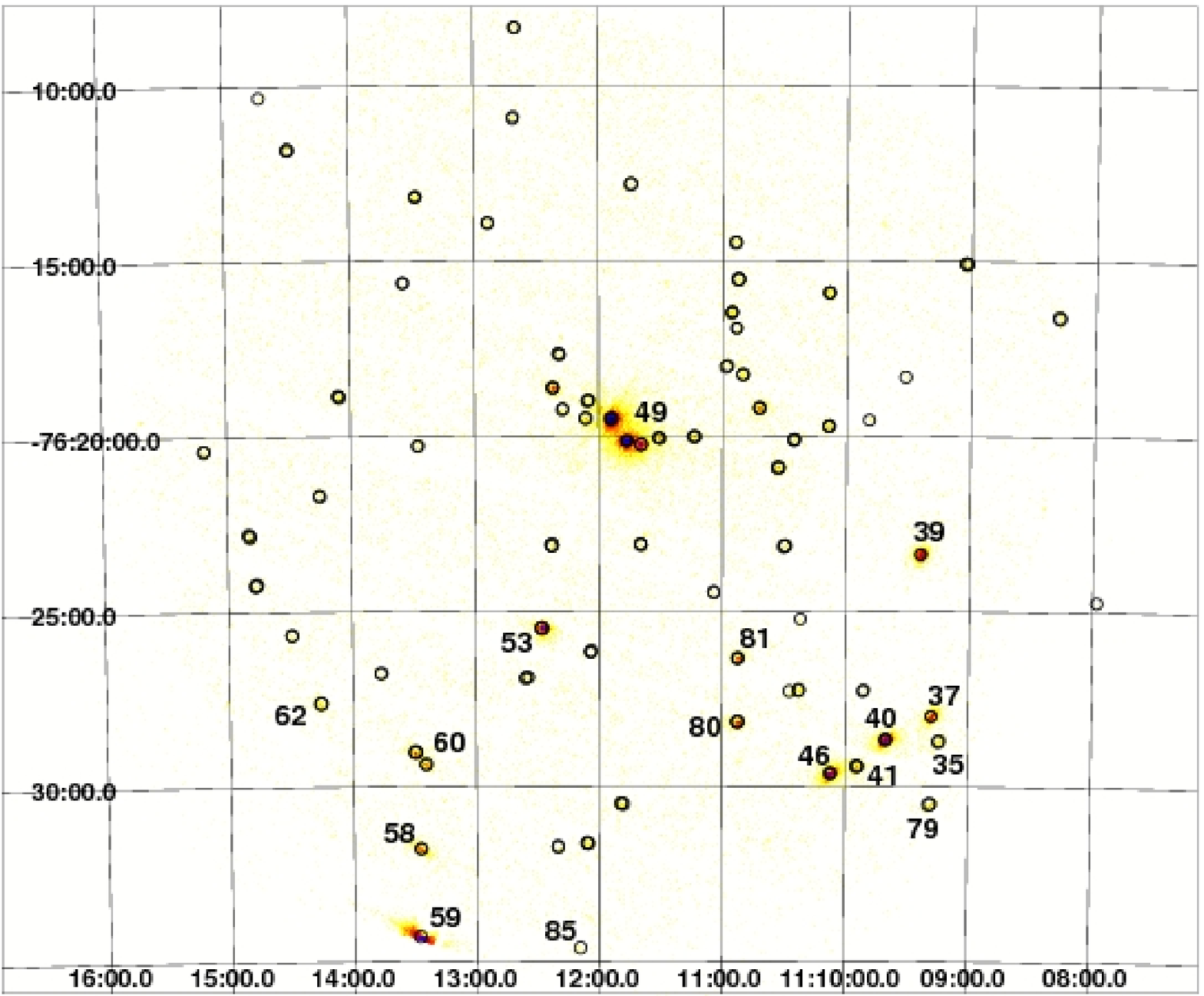}
      \caption{Image with all EPIC detectors combined and overplotted X-ray sources from the source detection runs. 
The already known CHXR sources are numbered.}
         \label{field}
   \end{figure}

The XMM-Newton field of view is full of new X-ray sources, not detected by ROSAT and previously unknown. We have run a detection algorithm 
on the images in the energy band 0.5\,--\,5.0\,keV for all three EPIC detectors.
Sources above a detection likelihood threshold of 30 
and multiple detections with a threshold of 20/15 (PN/MOS) are shown in Fig.\,\ref{field}. 
This translates to a minimum number of source counts of roughly 100 for the PN and 50 for each MOS detector and an observed X-ray luminosity of
$\sim 10^{28}$\,erg/s, assuming T~Tauri star like spectra. 
Each detection is indicated by a circle with 10\arcsec\, radius, 
given position uncertainties are in the range from 0.1\arcsec\, for the brighter targets up to
1--2\arcsec\, for faint sources, especially near the edge of the field.
However, inspection of the images and cross check of the
derived positions from different instruments indicates larger positional errors in the range 
of 1\arcsec for the brightest sources up to 3\arcsec for fainter sources at the edge of the field.
We checked for systematic position offsets using the brighter targets from Sect.\ref{reschxr} with well known positions, but
find no systematic deviations between measured and given positions. We note that the position accuracy from XMM-Newton is 
significantly higher than from ROSAT and uncertainties are comparable to deviations found in various references for optical/IR positions. 
Most sources are detected independently in all EPIC detectors and the determined positions agree very well, while
most double and the few single detections are exclusively due to
non overlapping fields of view, chip failure (CCD6 in MOS1), chip gaps or bad columns. 
This ensures that the given detections are with high confidence true X-ray sources.

We detect 71 X-ray sources, compared to 16 resolved ROSAT sources, increasing the number of X-ray sources in this region by a factor of four.
While some extragalactic sources are certainly present, many sources are likely cloud members. 
X-ray colours and infra-red emission are used to separate possible stellar objects from extragalactic sources like background AGN.
We derive exposure map weighted X-ray count rates in three energy bands (soft: 0.5\,--\,1.0\,keV, medium: 1.0\,--\,2.0\,keV, hard: 2.0\,--\,7.5\,keV) 
and perform a position cross check with the 2MASS point source catalog (\cite{2mass}). 
The positional accuracy of XMM-Newton and 2MASS is sufficient to clearly and unambiguously identify possible counterparts.
Errors on the derived count rates depend by a factor of up to two on the respective detector position, for brighter sources above 0.1\,cts/s
they are negligible, at $10^{-2}-10^{-3}$\,cts/s they are around 5--10\% and increase for faint sources with $5\times10^{-4}$\,cts/s
up to 30\%. Band fluxes below these value should be taken only as rough estimates. 
The count rate to energy conversion factors depend of course on the assumed spectra. 
We use the properties derived in Sect.\,\ref{reschxr} and derive
count rate to flux factors for PN data in the energy band 0.5\,--\,7.5\,keV. For unabsorbed stellar sources (CHXR~53)
they are around $1.5\times 10^{-12}$\,erg\,cts$^{-1}$\,cm$^{-2}$, they increase for 
more absorbed or hotter stellar targets like Hn\,15\,/\,CHXR\,79 to $2.0/2.5\times 10^{-12}$\,erg\,cts$^{-1}$\,cm$^{-2}$
and reach for hard sources (CHXR~80) values around $3.5\times 10^{-12}$\,erg\,cts$^{-1}$\,cm$^{-2}$. 
The MOS count rates are roughly a factor three lower,
using bright sources we find an average ratio of 3.2 for the total count rate (4.1/2.6/2.7 for the soft/medium/hard energy band).

\begin{table*}[!ht]
\setlength\tabcolsep{5pt}
\caption{\label{list}X-ray source list. Count rates are given
per instrument (0.5-7.5\,keV in $10^{-1}$\,cts/s and in three energy bands 0.5\,--\,1.0, 1.0\,--\,2.0, 2.0\,--\,7.5\,keV).
Sources that are not present in a detector are marked by dashes, sources without useful data by the detector ID in brackets.}
{\scriptsize
\begin{tabular}{lllllllll}\hline\hline
CHXX &RA (h, m, s) & Dec (deg, \arcmin, \arcsec) &Err.(\arcsec) & PN & MOS1 & MOS2 & 2MASS/DENIS-ID$^1$ &Other$^2$ \\\hline
1  & 11\, 07\, 58.1    & -76\, 24\, 43.5  & 3.  &--& --& 0.043 - 0.5/1.8/2.1e-3 & -- & \\
2  & 11\, 08\, 17.6    & -76\, 16\, 35.5  & 3.  &(P) & 0.040 - 2.0/1.8/0.2e-3 & 0.027 - 1.2/1.4/0.1e-3& 2M11081745-7616346 & U0137-0067981\\
3  & 11\, 09\, 02.6    & -76\, 15\, 04.0  & 2.  &0.066 - 3.2/3.1/0.3e-3&0.012 - 5.1/7.2/0.0e-4 & 0.023 - 0.8/1.3/0.2e-3 & 2M11090277-7615034 & U0137-0068083\\
4  & 11\, 09\, 14.2    & -76\, 28\, 41.0  & 2.  &0.16~~ - 8.5/5.9/1.2e-3& -- & 0.048 - 2.1/1.9/0.9e-3& 2M11091380-7628396 & CHXR\,35, Hn 8\\
5  & 11\, 09\, 18.0    & -76\, 27\, 59.0  & 2.  &1.21~~ - 5.5/5.2/1.4e-2& -- & 0.37~~ - 1.3/1.8/0.5e-2& 2M11091769-7627578 & CHXR\,37, CCE1-74\\
6  & 11\, 09\, 18.4    & -76\, 30\, 30.0  & 2.  &0.22~~ - 0.2/1.2/0.8e-2&-- & 0.11~~ - 0.4/5.3/5.0e-3 & 2M11091812-7630292 & CHXR\,79, Hn 9\\
7  & 11\, 09\, 23.7    & -76\, 23\, 22.0  & 2.  &1.07~~ - 3.4/4.5/2.8e-2&-- & 0.36~~ - 0.8/1.7/1.1e-2 & 2M11092379-7623207 & CHXR\,39, VZ Cha\\
8  & 11\, 09\, 31.7    & -76\, 18\, 18.5  & 3.  &(P)& -- & 0.012 - 1.5/5.0/5.5e-4& -- & \\
9  & 11\, 09\, 40.3    & -76\, 28\, 40.0  & 1.  &1.52~~ - 7.8/6.0/1.4e-2&0.51~~ - 2.0/2.5/0.6e-1 & 0.46~~ - 1.8/2.2/0.6e-2& 2M11094006-7628391 & CHXR\,40, CCE1-78\\
10 & 11\, 09\, 49.3    & -76\, 19\, 32.5  & 3.  &(P)&-- & 0.014 - 0.4/4.7/8.7e-4 & -- &\\
11 & 11\, 09\, 51.3    & -76\, 27\, 16.0  & 2.  &0.025 - 0.6/0.8/1.1e-3&-- & 0.013 - 0.9/6.1/5.7e-4 & -- &\\
12 & 11\, 09\, 54.1    & -76\, 29\, 26.0  & 2.  &0.24~~ - 0.8/1.3/0.4e-2& 0.081 - 1.9/4.3/1.9e-3 & 0.086 - 2.0/4.7/2.0e-3& 2M11095407-7629253 &CHXR\,41, Sz 33\\
13 & 11\, 10\, 07.1    & -76\, 29\, 38.5  & 1.  &1.46~~ - 6.4/6.4/1.8e-2& 0.47~~ - 1.6/2.4/0.7e-2 & 0.51~~ - 1.7/2.6/0.8e-2& 2M11100704-7629376 & CHXR\,46, WY Cha\\
14 & 11\, 10\, 08.5    & -76\, 15\, 55.0  & 1.  &0.074 - 1.6/3.1/2.7e-3& 0.022 - 5.8/9.3/7.8e-4 & 0.024 - 0.3/1.3/0.9e-3& -- &\\
15 & 11\, 10\, 08.8    & -76\, 19\, 43.0  & 2.  &0.073 - 0.3/2.6/4.4e-3& 0.027 - 0.0/0.9/1.8e-3 & 0.035 - 0.0/1.1/2.4e-3& -- &\\
16 & 11\, 10\, 22.2    & -76\, 25\, 14.0  & 3.  &(P)& -- & 0.012 - 7.1/3.2/1.7e-4& 2M11102226-7625138 & ChaI 710 (BD)\\
17 & 11\, 10\, 22.7    & -76\, 27\, 16.0  & 2.  &0.096 - 1.9/4.1/3.5e-3&-- & 0.041 - 0.5/1.6/2.1e-3 & -- & \\
18 & 11\, 10\, 25.4    & -76\, 20\, 07.0  & 2.  &0.028 - 0.9/1.0/0.9e-3& 0.011 - 2.4/4.8/3.3e-4 & 0.008 - 0.9/4.4/2.8e-4& -- & \\
19 & 11\, 10\, 27.4    & -76\, 27\, 17.5  & 3.  &(P)& -- & 0.011 - 0.1/3.8/7.4e-4& -- &  \\
20 & 11\, 10\, 29.8    & -76\, 23\, 09.0  & 2.  &0.027 - 1.1/1.0/0.5e-3& 0.011 - 3.2/4.4/3.1e-4 & 0.007 - 1.3/3.6/2.1e-4& -- &\\
21 & 11\, 10\, 33.1    & -76\, 20\, 55.0  & 1.  &0.063 - 1.9/2.3/2.1e-3& 0.018 - 6.1/8.2/3.5e-4 & 0.023 - 0.4/1.1/0.8e-3& --& \\
22 & 11\, 10\, 42.2    & -76\, 19\, 13.0  & 1.  &0.14~~ - 4.0/5.8/3.8e-3&0.050 - 1.0/2.4/1.6e-3 & 0.051 - 1.2/2.1/1.8e-3 &-- & \\
23 & 11\, 10\, 50.2    & -76\, 18\, 15.5  & 2.  &(P)&0.029 - 1.5/1.2/0.2e-3 & 0.032 - 1.5/1.3/0.4e-3 & 2M11105039-7618145 &  U0136-0067477\\
24 & 11\, 10\, 52.4    & -76\, 15\, 32.0  & 1.  &0.049 - 0.0/1.4/3.6e-3& 0.011 - 0.3/2.9/7.5e-4 & 0.013 - 0.0/0.2/1.1e-3& --&  \\
25 & 11\, 10\, 52.5    & -76\, 26\, 21.5  & 1.  &0.18~~ - 4.6/7.3/6.0e-3&0.065 - 1.1/3.0/2.4e-3 & 0.061 - 1.3/2.6/2.2e-3 & 2M11105228-7626223 & CHXR\,81\\
26 & 11\, 10\, 52.7    & -76\, 28\, 10.5  & 1.  & 0.38~~ - 0.7/1.5/1.7e-2& 0.13~~ - 2.0/5.6/5.1e-3 & 0.11~~ - 1.7/5.1/4.5e-3& -- & CHXR\,80 \\
27 & 11\, 10\, 53.2    & -76\, 16\, 56.0  & 2.  &(P)&0.008 - 3.4/4.5/0.3e-4 & 0.009 - 2.4/5.8/1.1e-4 & 2M11105331-7616559 & U0137-0068316 \\
28 & 11\, 10\, 53.7    & -76\, 14\, 29.0  & 2.  &(P)& 0.020 - 3.7/8.9/7.5e-4 & 0.018 - 0.3/1.1/0.5e-3& -- &  U0137-0068321\\
29 & 11\, 10\, 55.6    & -76\, 16\, 29.0  & 1.  & 0.036 - 0.2/1.7/1.6e-3& 0.013 - 0.9/5.4/6.7e-4 & 0.016 - 2.0/6.8/7.3e-4& -- &  \\
30 & 11\, 10\, 57.9    & -76\, 18\, 00.0  & 3.  &(P)&0.011 - 3.1/5.2/2.9e-4 & 0.012 - 3.1/5.5/2.9e-4& 2M11105830-7617580  & U0137-0068332 \\
31 & 11\, 11\, 04.2    & -76\, 24\, 28.5  & 3.  &(P)&0.008 - 0.0/2.0/5.8e-4 & 0.008 - 0.0/2.2/5.7e-4 & -- & \\
32 & 11\, 11\, 13.8    & -76\, 20\, 02.0  & 2.  &0.034 - 1.0/1.4/1.1e-3& 0.012 - 1.2/7.2/3.8e-4 & 0.017 - 3.8/6.9/6.1e-4& -- & \\
33 & 11\, 11\, 30.7    & -76\, 20\, 03.5  & 2.  &0.068 - 4.1/2.6/0.3e-3& 0.022 - 1.0/1.0/0.1e-3 & 0.020 - 0.9/1.1/0.0e-3& 2M11113049-7620030 &  U0136-0067600\\
34 & 11\, 11\, 39.6    & -76\, 23\, 07.0  & 2.  &0.020 - 4.7/6.1/8.8e-4& 0.011 - 1.4/6.8/2.3e-4 & 0.007 - 1.9/4.1/1.5e-4& --& \\
35 & 11\, 11\, 39.7    & -76\, 20\, 15.5  & 1.  &0.59~~ - 2.3/2.7/0.9e-2& 0.19~~ - 0.6/1.0/0.3e-2 & 0.19~~ - 0.6/1.0/0.3e-2& 2M11113965-7620152 & XX Cha\\
36 & 11\, 11\, 44.2    & -76\, 12\, 49.0  & 2.  &(P)& 0.018 - 5.1/5.0/8.0e-4 & 0.013 - 2.3/4.7/5.5e-4& --&  \\
37 & 11\, 11\, 46.4    & -76\, 20\, 09.0  & 1.  & 2.51~~ - 1.1/1.1/0.4e-1& 0.84~~ - 2.6/4.3/1.4e-2 & 0.79~~ - 2.4/4.2/1.3e-2& 2M11114632-7620092& CHX\,18N, CCE1-103\\
38 & 11\, 11\, 49.1    & -76\, 30\, 31.0  & 2.  & 0.074 - 1.8/3.1/2.5e-3& 0.020 - 0.2/1.2/0.7e-3 & 0.021 - 0.2/1.2/0.7e-3& -- & \\
39 & 11\, 11\, 54.0    & -76\, 19\, 31.0  & 1.  & 4.29~~ - 2.2/1.6/0.5e-1& 1.33~~ - 5.5/5.9/1.9e-2 & 1.36~~ - 5.2/6.5/2.0e-2& 2M11115400-7619311& Hn 15 \\
40 & 11\, 12\, 03.9    & -76\, 26\, 10.0  & 2.  &0.021 - 4.5/9.0/7.5e-4& 0.005 - 0.8/4.1/4.8e-4 & 0.012 - 0.8/3.4/7.5e-4 &--&  \\
41 & 11\, 12\, 05.3    & -76\, 19\, 00.5  & 2.  &(P)&0.013 - 0.2/5.8/7.1e-4 & 0.013 - 0.0/3.4/9.4e-4 & --& \\
42 & 11\, 12\, 05.6    & -76\, 31\, 38.0  & 2.  &0.13~~ - 2.3/5.8/4.5e-3& 0.046 - 0.2/2.1/2.3e-3 & 0.051 - 0.6/2.7/1.8e-3& --& \\
43 & 11\, 12\, 06.4    & -76\, 19\, 31.0  & 2.  &(P)& 0.019 - 4.9/7.4/6.7e-4 & 0.015 - 3.3/8.6/3.2e-4& --& U0136-0067713\\
44 & 11\, 12\, 09.6    & -76\, 34\, 37.0  & 3.  &0.061 - 1.8/3.3/0.9e-3&(M1) & -- & 2M11120984-7634366 & CHXR\,85, Sz 40\\
45 & 11\, 12\, 17.5    & -76\, 19\, 16.0  & 2.  &(P)& 0.014 - 3.3/6.4/3.9e-4 & 0.012 - 1.4/7.8/3.2e-4& --& \\
46 & 11\, 12\, 19.1    & -76\, 17\, 40.5  & 2.  &0.020 - 1.1/0.9/0.0e-3& 0.006 - 2.1/2.4/1.1e-4 & 0.008 - 2.0/5.2/0.6e-4& 2M11121884-7617400 &  U0137-0068518\\
47 & 11\, 12\, 20.1    & -76\, 31\, 44.0  & 3.  &0.048 - 1.0/1.6/2.2e-3& 0.015 - 2.6/5.7/6.2e-4 & 0.017 - 2.5/7.3/6.7e-4& -- &  \\
48 & 11\, 12\, 22.2    & -76\, 18\, 38.5  & 1.  &0.16~~ - 4.0/6.8/5.3e-3&0.047 - 1.0/2.0/1.7e-3 & 0.046 - 1.1/1.9/1.6e-3 & --& \\
49 & 11\, 12\, 22.9    & -76\, 23\, 08.0  & 1.  &0.043 - 0.8/2.1/1.3e-3&0.013 - 1.4/6.6/5.1e-4 & 0.014 - 1.0/5.9/7.0e-4 & --&  \\
50 & 11\, 12\, 27.8    & -76\, 25\, 29.5  & 1.  &0.83~~ - 5.9/2.1/0.3e-2& 0.24~~ - 1.5/0.8/0.1e-2 & 0.26~~ - 1.6/0.9/0.1e-2& 2M11122775-7625293& CHXR\,53, CCE1-117\\
51 & 11\, 12\, 34.8    & -76\, 26\, 54.0  & 2.  &0.058 - 1.3/1.8/2.7e-3& 0.018 - 3.1/9.3/5.2e-4 & 0.016 - 0.2/1.0/0.4e-3& -- & \\
52 & 11\, 12\, 40.3    & -76\, 08\, 20.5  & 2.  &0.024 - 1.3/0.7/0.4e-3& 0.014 - 5.2/5.7/3.5e-4 & 0.011 - 3.5/5.9/1.7e-4& 2M11124044-7608195 & U0138-0069520 \\
53 & 11\, 12\, 41.2    & -76\, 10\, 55.5  & 2.  &0.032 - 0.2/1.5/1.6e-3& 0.008 - 0.0/5.7/2.2e-4 & 0.013 - 1.1/4.6/7.1e-4&--& \\
54 & 11\, 12\, 53.0    & -76\, 13\, 55.0  & 2.  &0.026 - 0.3/1.0/1.3e-3& 0.008 - 0.4/1.4/6.6e-4 & -- & --& \\
55 & 11\, 13\, 24.6    & -76\, 29\, 22.5  & 2.  &0.24~~ - 1.4/0.8/0.1e-2& 0.064 - 3.1/3.0/0.4e-3 & 0.062 - 3.2/2.4/0.6e-3& 2M11132446-7629227& CHXR\,60SW, Hn 18\\
56 & 11\, 13\, 27.1    & -76\, 31\, 46.0  & 2.  &0.63~~ - 1.4/2.7/2.2e-2& 0.25~~ - 0.4/1.1/0.9e-2 & 0.21~~ - 3.3/9.9/8.2e-3 & D111327.02-763147.0& CHXR 58\\
57 & 11\, 13\, 27.5    & -76\, 20\, 17.0  & 2.  &(P)& 0.024 - 1.4/0.8/0.1e-3 & 0.023 - 1.3/0.8/0.2e-3& 2M11132747-7620175 & U0136-0067981\\
58 & 11\, 13\, 27.8    & -76\, 34\, 17.0  & 2.  &11.4~~ - 4.7/4.5/2.1e-1& -- &  --& 2M11132737-7634165 & CHXR\,59, CCE1-117\\
59 & 11\, 13\, 28.1    & -76\, 13\, 10.5  & 2.  &0.069 - 0.1/2.5/4.2e-3& 0.022 - 0.1/0.8/1.4e-3 & 0.025 - 0.1/0.8/1.6e-3& --& \\
60 & 11\, 13\, 29.6    & -76\, 29\, 01.0  & 1.  &0.38~~ - 2.1/1.5/0.2e-2&0.11~~ - 5.3/4.9/1.0e-3 & 0.11~~ - 5.9/4.1/0.9e-3 & 2M11132970-7629012&CHXR\,60NE,  Hn 19\\
61 & 11\, 13\, 34.2    & -76\, 15\, 38.0  & 2.  &(P)&0.010 - 0.8/4.8/4.8e-4 & 0.011 - 2.3/5.5/3.5e-4 & --&  \\
62 & 11\, 13\, 46.2    & -76\, 26\, 46.0  & 2.  &0.030 - 0.5/1.0/1.5e-3& -- & 0.013 - 0.4/6.2/6.3e-4& --&  \\
63 & 11\, 14\, 05.6    & -76\, 18\, 51.5  & 2.  &0.052 - 1.4/1.7/2.1e-3& 0.019 - 2.8/6.3/9.4e-4 & 0.016 - 3.0/5.1/8.0e-4& --& \\
64 & 11\, 14\, 15.2    & -76\, 21\, 41.5  & 2.  &0.029 - 0.6/1.4/0.9e-3& 0.010 - 1.0/6.6/2.6e-4 & 0.007 - 1.3/4.9/1.0e-4& --& \\
65 & 11\, 14\, 15.5    & -76\, 27\, 37.0  & 2.  &0.18~~ - 1.1/0.6/0.1e-2& 0.070 - 3.0/3.1/0.8e-3 & 0.055 - 2.1/2.8/0.5e-3& 2M11141565-7627364& CHXR\,62, Hn 20\\
66 & 11\, 14\, 29.0    & -76\, 11\, 48.0  & 3.  &0.060 - 1.1/2.8/2.1e-3& 0.020 - 0.3/1.0/0.7e-3 & 0.014 - 0.9/8.9/4.5e-4 &--& U0138-0069824 \\
67 & 11\, 14\, 29.1    & -76\, 25\, 40.0  & 2.  &0.041 - 3.1/1.0/0.0e-3& 0.014 - 8.0/3.1/2.5e-4 & 0.009 - 6.2/2.2/0.9e-4& 2M11142906-7625399 &U0135-0067114\\
58 & 11\, 14\, 42.5    & -76\, 10\, 19.0  & 3.  &0.056 - 1.5/2.4/1.8e-3& -- & --& --& \\
69 & 11\, 14\, 46.5    & -76\, 24\, 14.0  & 2.  &0.051 - 1.1/1.8/2.3e-3& 0.021 - 0.2/1.1/0.8e-3 & 0.018 - 0.0/0.8/1.1e-3& --& \\
71 & 11\, 14\, 49.3    & -76\, 22\, 49.0  & 2.  &0.078 - 1.9/3.7/2.2e-3& 0.024 - 0.5/1.1/0.9e-3 & 0.029 - 0.2/1.3/1.5e-3& --& \\
71 & 11\, 15\, 11.0    & -76\, 20\, 24.0  & 3.  &0.033 - 0.8/1.5/1.0e-3& -- & 0.020 - 2.8/8.1/8.8e-4& --&\\\hline
\end{tabular}

\begin{list}{}{}
\item 1: 2M for 2MASS\,J, D for DENIS database 3rd release; 
2: CHXR from \cite{fei93}, Hn from \cite{har93}, Sz from \cite{sch77}, CCE from \cite{cam98}, ChaI from \cite{lop04} and U for USNO-B1.0 catalog.
\end{list}
} 
\end{table*}

The derived count rates in the energy bands can be used as an indicator of the underlying type of source. 
Weakly or unabsorbed stellar sources are strongest in the soft band, moderate in the medium band and low in the hard band;
more absorbed sources loose mainly photons in the soft band and the medium band begins to dominate, but no known stellar source
from our sample peaks in the hard band. This holds even
in the most extreme cases in our sample, e.g. very strong absorption and a hot corona like
in CHXR~79, which is the CTTS closest to the cloud core, or strong flaring, which is present for CHXR~59.

A comprehensive list of all X-sources detected in this observation is given in Table\,\ref{list}. In the first column we introduce 
acronym CHXX (Chamaeleon X-ray XMM) to simplify further discussion, followed by the
source coordinates in the J2000 system and positional error (columns 2--4). The measured count rates are given for each instruments 
whenever available (columns 5--7), followed by proposed 
infra-red identifications with 2MASS or DENIS and likely optical counterparts (columns 8 and 9). 

The identifications of infra-red counterparts with 2MASS and DENIS succeeded only in roughly half of the cases.
Most IR-sources are detected in both surveys, in these cases we give the 2MASS identification.
The non-detections at infra-red wavelength are not strictly related to X-ray brightness and
some of the brighter sources, e.g. CHXR~80, exhibit no known infra-red or optical counterparts.
However, the X-ray sources not detected by 2MASS are predominantly hard X-ray sources and/or are quite faint, 
i.e with count rates \hbox{$\lesssim$\,$5\times10^{-3}$\,cts/s} per MOS detector.
Clearly most X-ray bright sources are young stars belonging to the cloud, but also some of the fainter X-ray sources
have 2MASS/DENIS and even possible stellar counterparts.
All known young stellar objects exhibit soft to medium spectral hardness in X-rays and
are also seen in 2MASS images, independently of their X-ray brightness. 
Further on, all other X-ray sources with soft stellar-like spectral hardness are present in IR/optical catalogues, supporting their nature 
as YSOs belonging to the cloud.
Beside 2MASS further identifications from SIMBAD are given. The soft source CHXX~16 is associated with 
the brown dwarf ChaI\,710 of spectral type M9 \cite{lop04},
while CHXX~27, 30 and 67 are probably associated with the young PMS stars GK\,38, GK\,40 and ESO\,H$\alpha$\,571,
see \cite{gom01} and \cite{com04}. 
On the other hand, sources with a hard flux distribution, i.e where the measured count rate increases from the soft to the hard band,
contribute with $\sim$\,15\% to our sample and do not posses IR counterparts in any case. Many of these sources, e.g. CHXX~15 and 59
are even much harder than the bright source CHXR~80 and probably of extragalactic origin.
The remaining unidentified X-ray sources exhibit intermediate spectral distributions, comparable to stronger absorbed stellar sources.
Contrary to the hard population, they are concentrated towards the dark cloud and known groups of T Tauri stars, indicating a
predominantly stellar origin.

\section{Summary and conclusions}
\label{sum}
 
  \begin{enumerate}
\item The XMM-Newton observation of the Cha I dark cloud revealed numerous X-ray sources. 
Observed X-ray luminosities of known T~Tauri stars in their quiescent phases span a broad range 
from log $L_{\rm X}$ of 28.4\,erg/s, up to 30.2\,erg/s,
whereas especially the classical T~Tauri stars are subject to strong absorption.
X-ray and bolometric luminosities are correlated in T Tauri stars, the average log $L_{\rm X}/L_{\rm bol}= -3.2$ is near the
saturation limit for magnetic activity.
We are able to spatially resolve several sources detected by ROSAT, making their stellar
counterparts unambiguous and confirm various possible detections. 
Additional new detections present in XMM-Newton images increase the number of known X-ray sources 
in this region to 71 sources, i.e by a factor of
roughly four. However, (stellar) counterparts for these sources are often unknown, but spectral properties, association
to the dark cloud and the available
IR/optical identifications indicate that a large portion is indeed of stellar origin.

\item We use medium resolution spectra of the brighter targets to determine spectral properties of the T~Tauri star population.
No significant differences are found in the spectral properties of classical and weak-line T~Tauri stars, whereas most stars
are clearly dominated by magnetic activity. This is reflected in emission measure distributions with most plasma residing at temperatures
of 8\,MK up to 25\,MK, whereas the X-ray brighter targets tend to have hotter coronae and a higher $L_{\rm X}/L_{\rm bol}$ ratio.
Accretional stars that are dominated by cool plasma are rare or absent; VZ~Cha is a possible case, but quite strong
absorption prevents a definite conclusion. Except the stronger average absorptions due to circumstellar material, 
the classical T~Tauri stars differ only by a lower  
$L_{\rm X}/L_{\rm bol}$ ratio compared to the weak line T~Tauri stars.

\item  Several flares occurred during our observation, additionally minor variability is present in several stars. 
The largest flare is
observed on the WTTS CHXR~59. It shows an increase in count rate of about 30 and typical coronal flare parameters, 
quite similar to young main sequence stars. Also the stronger variability observed in CTTS clearly points to a coronal origin
due to the dominance of hot plasma.
Comparison with the ROSAT measurements shows that the level 
of quiescent X-ray luminosity is fairly constant and long term variability exceeding factor two is very rare in these stars.  
 \end{enumerate}

\begin{acknowledgements}
This work is based on observations obtained with XMM-Newton, an ESA science
mission with instruments and contributions directly funded by ESA Member
States and the USA (NASA).
This research has made use of the SIMBAD database,
operated at CDS, Strasbourg, France.\\
J.R. acknowledges support from DLR under 50OR0105.
\end{acknowledgements}

\end{document}